\documentclass[a4paper,twocolumn,floatfix,showpacs,showkeys,amsmath,amssymb,nobibnotes,altaffilletter]{revtex4-1}
\usepackage{graphicx}
\usepackage{pslatex}
\usepackage{xspace}
\usepackage{subfigure}
\usepackage{multirow}
\usepackage{booktabs}
\usepackage{textcomp}
\usepackage[pdfauthor={Bjoern Gieseking},%
pdftitle={Excitation Dynamics in Low Band Gap Donor--Acceptor Copolymers and Blends},
pagebackref=false,
pdfkeywords={Excitation Dynamics, Copolymers, Charge Transfer, Bulk-Heterojunctions, Organic Solar Cells},
pdftex]{hyperref}
\hypersetup{
    colorlinks,%
    citecolor=black,%
    filecolor=black,%
    linkcolor=black,%
    urlcolor=black
}

\newcommand{\bfig}{\begin{figure}}
\newcommand{\efig}{\end{figure}}
\newcommand{\bit}{\begin{itemize}}
\newcommand{\eit}{\end{itemize}}

\graphicspath{{figs/}{bilder/pdf/}}

\begin{document}

\preprint{}

\title{Excitation Dynamics in Low Band Gap Donor--Acceptor Copolymers and Blends} 

\author{B.~Gieseking$^1$}\email{bjoern.gieseking@physik.uni-wuerzburg.de}
\author{B.~Jaeck$^{1}$}
\author{E.~Preis$^{2}$}
\author{S.~Jung$^{2}$}
\author{M.~Forster$^{2}$}
\author{U.~Scherf$^{2}$}
\author{C.~Deibel$^{1}$}
\author{V.~Dyakonov$^{1,3}$}

\affiliation{$^1$Experimental Physics VI, Julius-Maximilian University of W{\"u}rzburg, 97074 W{\"u}rzburg, Germany}
\affiliation{$^2$Macromolecular Chemistry and Institute for Polymer Technology, Bergische University of Wuppertal, D-42097 Wuppertal}
\affiliation{$^3$Bavarian Center for Applied Energy Research, ZAE Bayern, 97074 W{\"u}rzburg, Germany}

\date{\today}

\begin{abstract}
Donor--acceptor (D--A) type copolymers show great potential for the application in the active layer of organic solar cells. Nevertheless the nature of the excited states, the coupling mechanism between them and the relaxation pathways following photoexcitation are yet to be clarified. We carried out comparative measurements of the steady state absorption and photoluminescence (PL) on the copolymer poly[N-(1-octylnonyl)-2,7-carbazole]-alt-5,5-[4$'$,7$'$-di(thien-2-yl)-2$'$,1$'$,3$'$-benzothiadiazole] (PCDTBT), its building blocks as well as on the newly synthesized N-(1-octylnonyl)-2,7-bis-[(5-phenyl)thien-2-yl)carbazole (BPT-carbazole), which represents the PCDTBT segments without the thiadiazole groups (Fig. 1). The high-energy absorption band (HEB) of PCDTBT was identified with absorption of carbazoles with adjacent thiphene rings while the low-energy band (LEB) originates instead from the charge transfer (CT) state delocalized over the aforementioned unit with adjacent benzothiadiazole group. Photoexcitation of the HEB is followed by internal relaxation (electron transfer towards thiadiazole) prior the radiative decay to the ground state. Adding $\mathrm{PC_{70}BM}$ results in the efficient PL quenching within the first 50 ps after excitation. From the PL excitation experiments no evidence for a direct electron transfer from the HEB of PCDTBT towards the fullerene acceptor was found, therefore the internal relaxation mechanisms within PCDTBT can be assumed to precede. Our findings indicate that effective coupling between copolymer building blocks governs the photovoltaic performance of the blends.             
\end{abstract}

\maketitle

 \subsection{INTRODUCTION}
One of the successful approaches for improving the performance of organic solar cells includes the synthesis of new materials used in the active  device layer. While the best acceptors so far are still the widely used fullerene derivatives, the efficiency of organic solar cells could be significantly improved reaching an efficiency above 8 \% \cite{deibel_2010,green_2011} by the synthesis of new and more complex copolymers with the record of 10.6 \% in a tandem architecture \cite{wong_2012} . These so-called third generation polymers have a reduced band gap and are designed to offer a more efficient light harvesting of the solar spectrum and higher open circuit voltages \cite{blouin_2008,heeger_2010}. A prominent and intensively studied group of these new materials are the donor--acceptor copolymers, for example fluorene- or carbazole-based polymers \cite{svensson_2003,zhang_2006,beaupre_2011}. Apart from the enhanced device performance that was achieved using these donor materials little is known about the elementary processes taking place after photoexcitation, which however might be of importance to fully exploit the potential of these becoming more and more complex systems.

In this work we report on the excitation dynamics in the donor-acceptor copolymer poly[N-(1-octylnonyl)-2,7-carbazole]-alt-5,5-[4',7'-di(thien-2-yl)-2',1',3'-benzothiadiazole] PCDTBT. This promising new acceptor, which was first synthesized by M. Leclerc and coworkers \cite{blouin_2007}, exhibits an internal quantum efficiency approaching unity and yields efficiencies of up to 7.2 \% in solar cell devices when blended with $\mathrm{PC_{70}BM}$ \cite{park_2009,steirer_2011,sunb_2011}. In contrast to more conventional and widely studied conjugated polymers such as poly(phenylene--vinylene) or polythiophene (polymers of second generation), the absorption spectra of donor--acceptor copolymers exhibit two distinct absorption bands. In the case of PCDTBT the maxima of these bands are located at approximately 400 and 560 nm, respectively. These two bands have been assigned to $\pi-\pi^{*}$ transitions into the first and second excited singlet state \cite{tong_2010} which were later found to exhibit a CT character \cite{banerji_2012}. Contrary to these results in a theoretical work on polyfluorene-based copolymer DiO-PFDTBT only the lower-energy band was associated with intrachain charge transfer \cite{jespersen_2004}.

Here we carry out a comparative study of absorption and photoluminescence spectra of PCDTBT and its building blocks, the carbazole and 4,7-di(thien-2-yl)-2,1,3-benzothiadiazole (DTBT) monomers, as well as the newly synthesized N-(1-octylnonyl)-2,7-bis-[(5-phenyl)thien-2-yl)carbazole (BPT-carbazole). The latter resembles a segment of the copolymer backbone without the thiadiazole units (see \ref{fig:Molekule}). The measurements enable us to identify the high-energy absorption band (HEB) of PCDTBT with the N-(1-octylnonyl)-2,7-di(thien-2-yl)carbazole (BTC) unit and the low-energy band (LEB) with a charge transfer (CT) band formed by BTC and adjacent benzothiadiazole. This finding is in contrast to Banerji et al., who interpreted their results as both bands having CT character \cite{banerji_2012}. The large difference in the solvatochromic shift of the photoluminescence of all studied benzothiadiazole containing molecules provides strong evidence in favor of a charge transfer character of the LEB, rather than an energy transfer character. For analyzing the excitation dynamics we applied picosecond time-resolved photoluminescence spectroscopy to the thin films of neat copolymer PCDTBT and 1:1 PCDTBT:$\mathrm{PC_{70}BM}$ blend. Combining the results of steady state and time-resolved measurement we are able to draw a detailed picture of the processes following the photoexcitation of the donor copolymers in a fullerene blend system used for photovoltaic applications.   

 \subsection{EXPERIMENTAL SECTION}
\begin{figure}[htp]
        \centering
        \includegraphics[width=8.5cm]{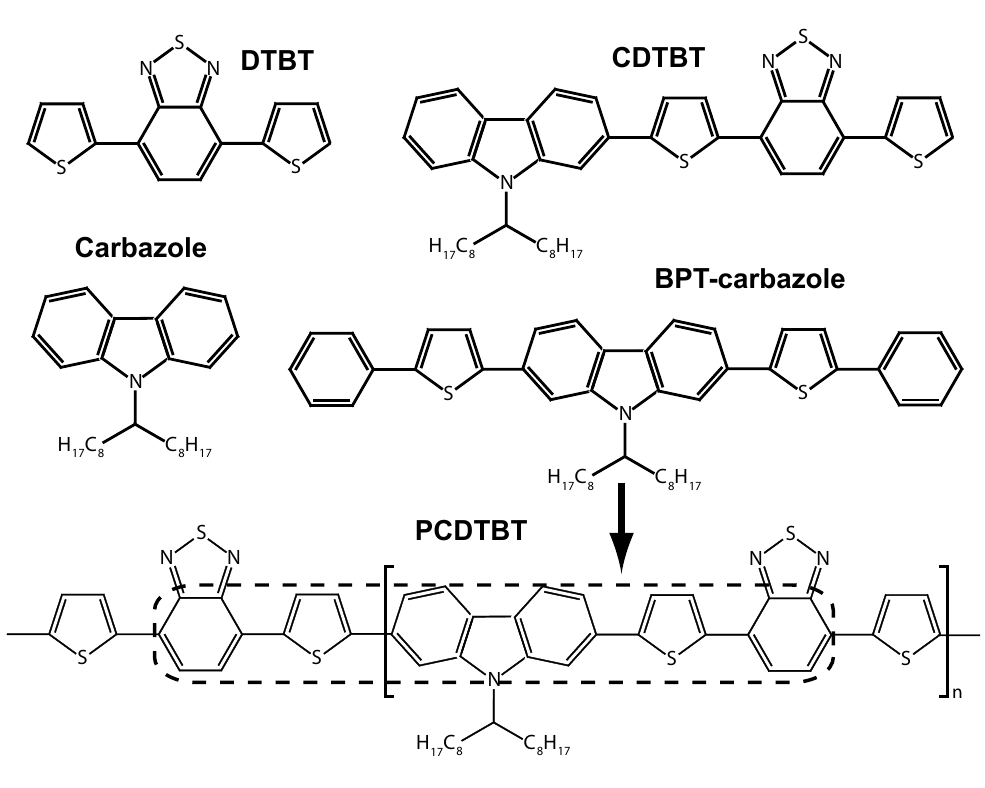}
        \caption{Molecular structures of the monomer DTBT, co-monomer CDTBT, monomer carbazole as well as BPT-carbazole investigated in this work. The BPT-carbazole resembles a segment of PCDTBT (at the bottom) without the thiadiazole groups as indicated by the dashed box.}
        \label{fig:Molekule}
\end{figure}

PCDTBT (1--material) and $\mathrm{PC_{70}BM}$ (Solenne) were used without further purification. All commercially available starting materials and reaction intermediates, reagents and solvents for the co-monomers and the monomer CDTBT (see Figure \ref{fig:Molekule}) were obtained from Acros, Aldrich or VWR and were used as supplied, unless otherwise noted. All syntheses were carried out in an inert atmosphere and the molecules were characterized by $^1$H- and $^{13}$C-NMR spectroscopy. The NMR spectra were recorded using a Bruker Avance 400 or a Bruker Avance III 600MHz spectrometer. The synthesis of the N-(1-octylnonyl)carbazole (carbazole) and 4,7-di(thien-2-yl)-2,1,3-benzothiadiazole (DTBT) were described elsewhere \cite{blouin_2007, zang_2004}. CDTBT was synthesized by a palladium-catalyzed Stille-type coupling using N-(1-octylnonyl)-2-bromocarbazole and 4-(thien-2-yl)-7-[2-(tributylstannyl)thien-5-yl]-2,1,3-benzothiadiazole \cite{vanmullekom_1998}. For the synthesis of BPT-carbazole, N-(1-octylnonyl)-2,7-dibromocarbazole (1.00 g; 1.77 mmol) and 5-phenyl-2-thienylboronic acid (0.84 g; 4.12 mmol) were dissolved in a mixture of 50 ml toluene, 10 ml water and 10 ml 2N aq. KOH solution. Tetrakis(tripehylphosphin)palladium was added and the resultant mixture was heated at 80°~ over night. The reaction mixture was allowed to cool, poured into water (300 ml) and the crude product was extracted with DCM (3x200 ml). The combined organic extracts were washed with brine (2x200 ml), dried over MgSO$_{4}$ and the solvent was removed under reduced pressure. Purification was carried out via column chromatography (hexane) to yield a yellow solid (340 mg; 29~\%).

For steady-state liquid phase optical measurements the materials were dissolved in chloroform except for the solvatochromism measurements, where cyclohexane and acetonitrile were additionally used as solvents (all spectroscopy grade). The solvent polarity function $f(D)-0.5f(n^{2})$ for the three solvents was calculated using $f(D)=(D-1)/(2D+1)$ and $f(n^{2})=(n^{2}-1)/(2n^{2}+1)$, where $n$ is the refractive index and $D$ the dielectric constant of the solvent~\cite{stahl_2006}. Thin films were fabricated by spincoating from chlorobenze solution onto a quartz substrate under nitrogen atmosphere. The film thickness of 90 nm was determined using a Dektak profilometer.

Absorption measurements were carried out using a 1 cm quartz cuvette on a JASCO V-670 UV/Vis/NIR spectrometer while for the PL and excitation measurements a PTI (Photon Technology International) QM-2000-4 fluorescence spectrometer with a cooled photomultiplier (R928 P) and a 75 W Xe short arc lamp (UXL-75XE, Ushio) was used. For the time-resolved PL measurements, the output of a Ti:Sa oszillator (Spectra Physics, 100 fs, 800 nm) was frequency doubled and focused onto the sample, which was mounted inside a liquid helium cryostat, using a fluence of $\mathrm{0.8~\mu W/cm^{2}}$. The PL was spectrally dispersed inside a spectrograph and detected with a C 5680 streak camera (Hamamatsu). The temporal resolution of the set-up described  was 8 ps. For XRD measurements the same thin film samples as for the optical measurements were investigated by a GE XRD3003T/T diffractometer (General Electric).

 \subsection{RESULTS AND DISCUSSION}
\subsubsection{COMPARATIVE STEADY STATE ABSORPTION AND PL MEASUREMENTS}
\begin{figure}[ht]
        \centering
        \includegraphics[width=9cm]{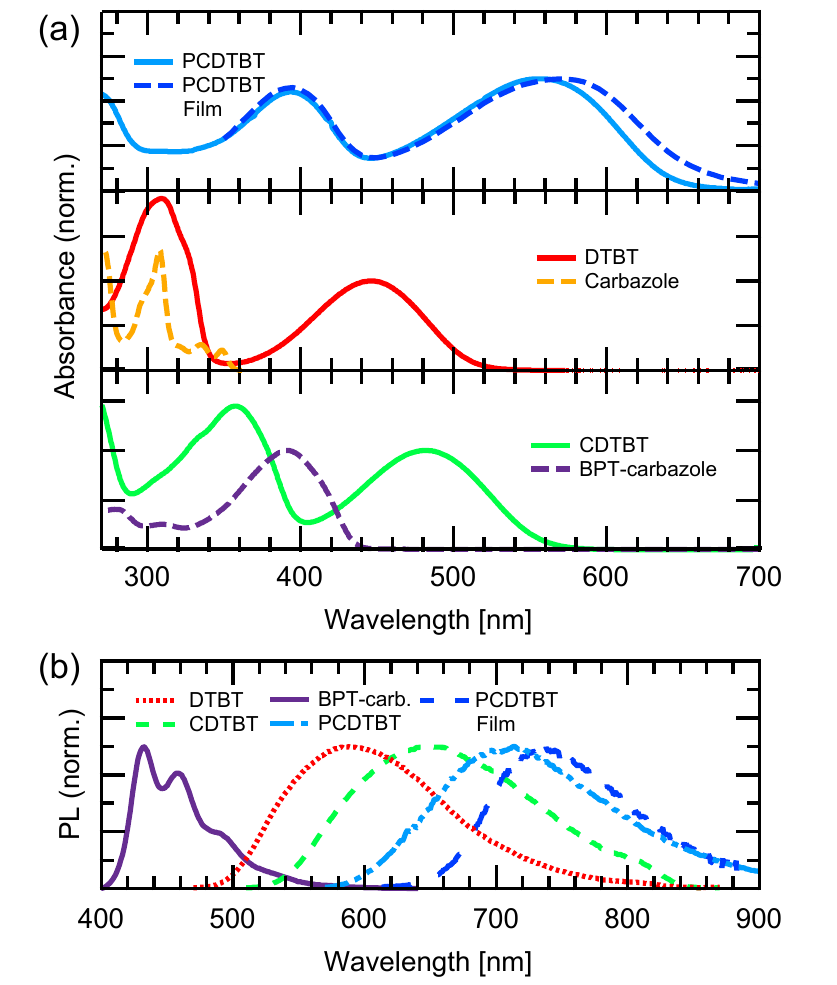}
        \caption{Absorption (a) and photoluminescence spectra (b) of the PCDTBT thin film and chloroform solution as well as of PCDTBT building blocks, DTBT, carbazole and co-monomer CDTBT, as well as BPT-carbazole in chloroform.}
        \label{fig:AbsPL}
\end{figure}

In order to identify the origin of the distinct absorption features in PCDTBT, we compared the absorption spectra of the PCDTBT copolymer, both in solution and in solid state, with the absorption of its building blocks---carbazole, DTBT and CDTBT, i.e., PCDTBT co-monomer unit---in solution. The results are shown in Figure \ref{fig:AbsPL} (a). Note, that the spectra presented, except for carbazole, are normalized with respect to the low energy peak height. Carbazole and DTBT exhibit absorption below 280 nm, which is the typical region for benzene absorption, \cite{inagaki_1972} and an absorption band at around 308 nm. DTBT exhibits a second absorption peak located at 446 nm, which is absent in the carbazole spectrum and has a characteristic shape observed in molecules containing benzothiadiazole units \cite{chappell_2003}.
Comparing the absorption spectra of DTBT with CDTBT, the most prominent feature observed is the 40 nm redshift, although the spectral shapes are nearly identical. This shift can be explained by the delocalization of the $\pi$-electrons due to the overlap of the electron wavefunctions of the constituing monomers DTBT and carbazole \cite{kirova_2010}. In contrast to the comparative absorption study by Banerji et al. \cite{banerji_2012} we observe a slight broadening of the bands compared to DTBT, possibly due to conformational isomerism of the investigated CDTBT molecules. Similarly to DTBT, the CDTBT spectrum exhibits two absorption peaks whereas the low-energy bands at around 440 nm and 480 nm have the same shape and, thus, can be assumed to be of the same origin, as also indicated by theoretical calculations of Heeger and coworkers \cite{banerji_2012}. Red shifted CDTBT exhibits an additional absorption peak, seen below 280 nm, and shoulder at 300 nm which may indeed stem from the carbazole. We note, however, that the resulting CDTBT spectrum can not be seen as a simple superposition of the carbazole and DTBT spectra. Polymerization of CDTBT to PCDTBT leads to a further redshift of the two prominent peaks of about 70 nm due to an increased delocalization of the $\pi$-electrons and a spectral broadening while Banerji et al. observe a slight narrowing. The increased energetic broadening found in our measurements might be explained by a conjugation length distribution in the copolymer, also known for other polymers such as P3HT \cite{Salaneck_1988,feng_2007}. The thin film absorption is very similar to the absorption of PCDTBT in solution. Only the LEB exhibits a small redshift and slight broadening indicating its sensitivity to interchain coupling effects \cite{clark_2007} in contrast to the HEB. 

The shape and spectral position of the photoluminescence spectrum of PCDTBT do not depend on the excitation wavelength, which implies some initial relaxation process prior to radiative relaxation to the ground state. According to our previous analysis of the absorption spectra, such an internal relaxation mechanism should exist in all investigated molecules containing benzothiadiazole units. As shown in Figure \ref{fig:AbsPL} (b) the spectral shapes of DTBT, CDTBT and PCDTBT PL are identical, with the position of the maximum shifted to longer wavelength possibly due to increased delocalization of the $\pi$-electrons but strongly indicating the common final emissive state. In agreement with our results from the absorption measurements the PL spectra become slightly broadened when extending the molecules from DTBT to PCDTBT.
\begin{figure}[ht]
        \centering
        \includegraphics[width=8.8cm]{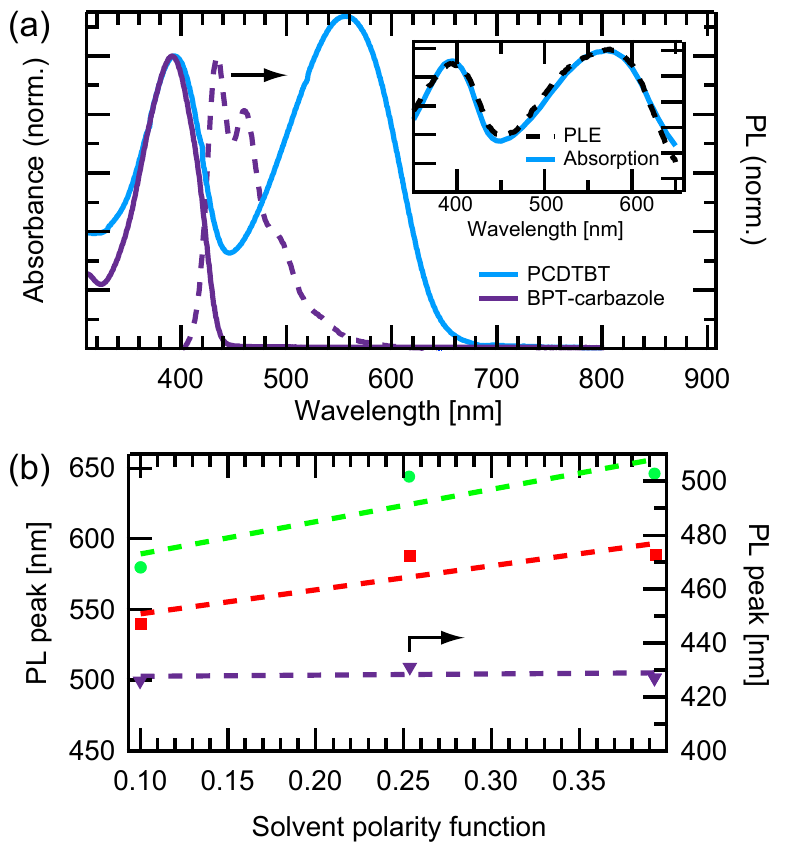}
        \caption{(a) Comparison of the absorption spectra  of BPT-carbazole and PCDTBT in solution showing the excellent overlap of the high-energy bands. Inset: The height of the absorption peaks of PCDTBT film can be exactly reproduced by its PLE spectrum indicating an efficient excitation transfer. (b) Photoluminescence peak position of CDTBT (circles), DTBT (squares) as well as BPT-carbazole (triangles) as a function of solvent polarity (cyclohexane, chloroform and acetonitrile).}
        \label{fig:Uberlapp}
\end{figure}

In order to identify the nature of the two prominent electronic states in PCDTBT, the coupling and hence the relaxation mechanism between them we carried out steady state absorption and PL measurements on newly synthesized BPT-carbazole (Figure \ref{fig:Molekule}). This molecule represents a sequence of the PCDTBT backbone with the phenyl instead of the thiadiazole groups. As shown in figure \ref{fig:Uberlapp} (a), the high-energy absorption band of PCDTBT is nearly identical with the absorption of BPT-carbazole.
By replacing the benzothiadiazole groups in PCDTBT by phenyl groups as in BPT-carbazole, the low energy band vanishes, indicating that this band is directly connected to the presence of benzothiadiazole units in the copolymer. The shape of the BPT-carbazole PL spectrum (Figure \ref{fig:Uberlapp} (a), dashed curve) is similar to that of polycarbazole \cite{morin_2002} (not shown), moreover, it is spectrally overlapping with the low-energy absorption band of PCDTBT, making the F\"orster-type coupling between the two energy states a plausible scenario.   

To verify this scenario, we investigated the influence of the solvent polarity on the photoluminescence peak position. Three different solvents with respective polarity function values of 0.10, 0.25 and 0.39 were used. Figure \ref{fig:Uberlapp} (b) shows the spectral positions of the PL maxima of co-monomer CDTBT (circles), monomer DTBT (squares) and BPT-carbazole (triangles) for the different solvents. The solvatochromic redshift of the PL peak position of the CDTBT with increasing solvent polarity is a strong indication for the charge transfer character of the LEB. This effect is also clearly present for benzothiadiazole containing building block DTBT, while it is absent in the case of thiadiazole-free BPT-carbazole. Therefore we conclude that the LEB of PCDTBT exhibits a CT character, which is consistent with recent theoretical and experimental results for a donor-acceptor molecule containing benzothiadiazole \cite{polander_2011} and the solvatochromic effect observed by Banerji et al. for both molecules \cite{banerji_2012}. In contrast to the latter work, where both bands of PCDTBT were associated with a charge transfer process, our experimental results on BPT-carbazole presented in Figure \ref{fig:Uberlapp} strongly indicate that the HEB of PCDTBT is a $\pi-\pi^{*}$ transitions without a significant CT character. The excitation is confined between two adjacent benzothiadiazole groups which hinder a further delocalization of the $\pi$-electrons. These findings are also able to explain the relatively small decrease in bandgap upon polymerization of CDTBT observed by Heeger and coworkers. The coupling between the HEB and the LEB accounting for the internal relaxation can be considered as a (partial) charge transfer from the BTC units towards the adjacent benzothiadiazole. As the shape of the absorption spectrum of the PCDTBT can be perfectly reproduced by its photoluminescence excitation (PLE) spectrum (see inset to Figure \ref{fig:Uberlapp} (a)) we consider the charge transfer as relevant decay mechanisms when exciting the polymer to the high energy electronic state.

\subsubsection{RELAXATION DYNAMICS IN PCDTBT AND PCDTBT:PC$_{70}$BM FILMS}
\begin{figure}[ht]
        \centering
        \includegraphics[width=8.5cm]{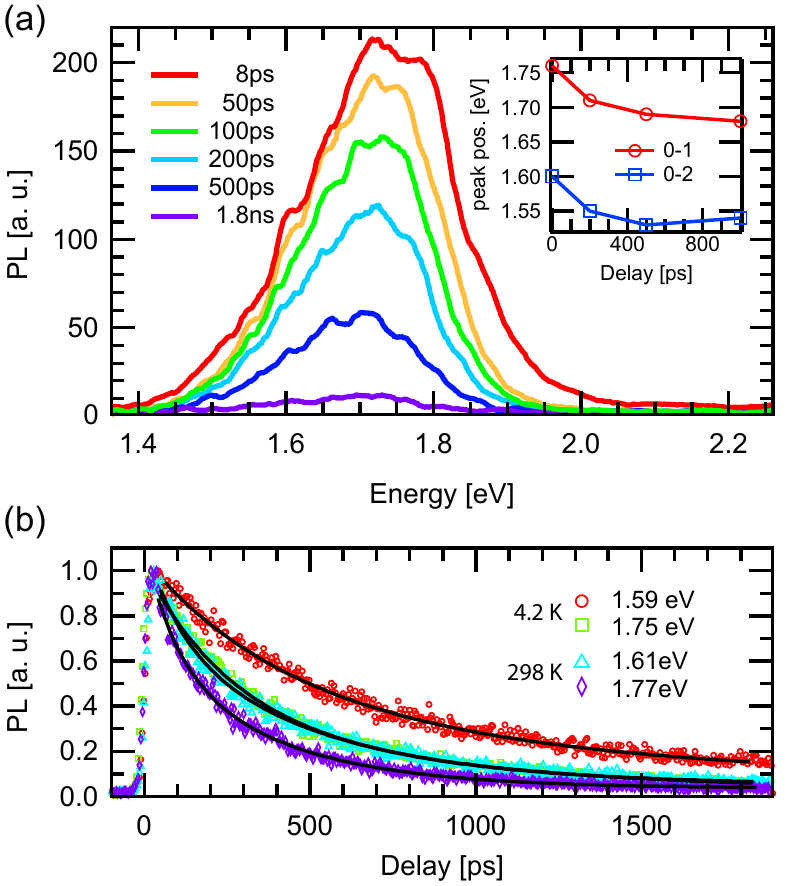}
        \caption{(a) PL spectra of PCDTBT for different delay times showing a dynamic redshift which can be ascribed to a shift of the 0-1 and 0-2 peak positions by approximately 70 meV within the first 500 ps (inset) due to exciton migration and conformational relaxation. (b) PL decay transients of PCDTBT thin film taken at the initial peak position (8 ps) of the dominant 0-1 and 0-2 transition at room temperature and 4.2 K.}
        \label{fig:TransSpecPCDTBT}
\end{figure}

The PL decay dynamics of the PCDTBT following excitation at 400 nm shown in Figure \ref{fig:TransSpecPCDTBT} (a) was found to exhibit a significant redshift. Similar to the corresponding steady state measurements, the PL spectra of PCDTBT thin film taken at various delay times can be described assuming a vibronic progression with the 0-1 and 0-2 transition being the dominant contributions to the spectrum (fits not shown). The dynamic shift of the peak positions of both transitions to lower energies for the first 500 ps (inset to Figure \ref{fig:TransSpecPCDTBT} (a)) reflects the shift of the PCDTBT PL spectrum, indicating that the spectral relaxation is not due to a change of the relative heights of the 0-1 and 0-2 transitions. The resulting total shift is around 70 meV, which is in good agreement with the shift observed for a similar PCDTBT type copolymer with longer side chains \cite{Laquai_2011}. It points to a reduced order of the copolymer, which is confirmed by X-ray diffraction measurements of PCDTBT thin film (not shown) suggesting only small crystallites of PCDTBT in a disordered surrounding. PL transients taken at the peak positions of the 0-1 and 0-2 transitions in the initial spectrum (8 ps) reflect this trend. The time constants of the low energy peak at 1.61 eV (see Table \ref{tab:PCDTBTTransients}), fitted by a bi-exponential decay model, are slightly larger compared to the transients measured at 1.77 eV. The average values for $\tau_{1}$ and $\tau_{2}$ of 98 and 425 ps, respectively, are in good agreement with the two long-lived PL decay components reported by Banerji et al. \cite{banerji_2010} where both components were assumed to originate from polymer backbone relaxation as well as exciton migration to lower energetic sites.
Comparing our time-resolved measurements at room temperature and at 4.2 K, we find that the amplitude ratio $\mathrm{A_{1}/A_{2}}$ of the two exponential contributions is changed by almost a factor of four. This can be explained by the reduced influence of exciton migration on the PL decay dyamics due to a decreased diffusivity at 4.2 K indicating that $\tau_{2}$ is mostly governed by this process. This assumption is consistent with the observed dynamic redshift of the PL spectrum within the first 500 ps at room temperature where $\tau_{2}$, i.e. the exciton migration to lower energetic sites, has a significant influence on the decay dynamics. Instead, the $\tau_{1}$ depicts the initial conformational relaxation of the polymer backbone accompanying the migration process. Note that for extraction of the low temperature transients the shift of the transitions to lower energies by 20 meV due to reduced thermal energy was taken into account.  
As we only observe emission from the low energy electronic state the internal relaxation within PCDTBT copolymer after photoexcitation has to be faster than the temporal resolution of our experiment of 8 ps, which is in accordance with Banerji et al., suggesting a sub-picosecond timescale \cite{banerji_2010}.
\begin{table}[ht]
	\centering
	\begin{tabular}{|c|c|c|c|c|}\hline \hline
	T~[K]&E~[eV]&$A_{1}/A_{2}$&$\tau_{1}$~[ps]&$\tau_{2}$~[ps]\\ \hline
	\multirow{2}{1cm}{298}& 1.61 & 0.38 & 104 & 480 \\ \cline{2-5}
	& 1.77 & 0.53 & 92 & 368 \\ \hline
	\multirow{2}{1cm}{4.2}& 1.59 & 1.47 & 420 & 1300 \\ \cline{2-5}
	& 1.75 & 1.93 & 267 & 910 \\ \hline \hline
	\end{tabular}
\caption{Parameters of the PL transients shown in Figure \ref{fig:TransSpecPCDTBT} (b) assuming bi-exponential decay. $E$ denotes the probe energy of the transients, $A_{1,2}$ and $\tau_{1,2}$ are the amplitudes and time constants, respectively.}
\label{tab:PCDTBTTransients}
\end{table}

\begin{figure}[ht]
	        \centering
	        \includegraphics[width=8.7cm]{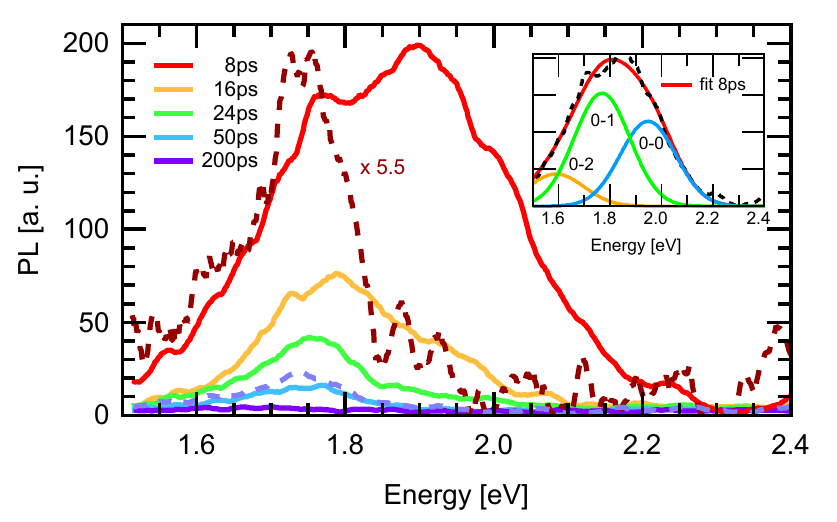}
	        \caption{PL specra of a 1:1 PCDTBT:$\mathrm{PC_{70}BM}$ blend at various delay times. PL spectra of neat $\mathrm{PC_{70}BM}$ 8 and 50 ps after excitation with 400 nm are shown for comparison (dashed lines). Inset shows the deconvolution of the initial (t=8 ps) PCDTBT:$\mathrm{PC_{70}BM}$ PL spectrum with 0-0, 0-1 and 0-2 vibronic transitions. The apparent initial high energy shift of the blend spectrum compared to PCDTBT spectrum (Fig. 4a) can be explained by an increased relative contribution of the 0-0 transition resulting in a spectral broadening of the copolymer PL spectrum.}
	        \label{fig:SpecBlend2}
\end{figure}

The PL of a 1:1 blend of PCDTBT with $\mathrm{PC_{70}BM}$ is strongly quenched due to efficient electron transfer to the fullerene resulting in a singlet exciton splitting (see Figure \ref{fig:SpecBlend2}). Assuming that $\mathrm{PC_{70}BM}$ acceptor has a nonzero absorption at the excitation wavelength of 400 nm a superposition of both PL spectra is expected. Comparing the spectral shape of transient spectra of the copolymer with separately recorded PL spectra of $\mathrm{PC_{70}BM}$ (dashed lines), we find that at 50 ps the blend spectrum can be completely described by acceptor emission, although the initial spectra differed significantly. This finding means that the copolymer PL signal is completely quenched within this time range, however the exact quenching rate cannot be determined within the temporal resolution of our setup. The blend PL spectrum is initially shifted to higher energies, as compared to neat PCDTBT PL, which can be explained by a dominant 0-0 vibronic transition (see inset to Figure \ref{fig:SpecBlend2}). This shift is due to reduced self-absorption of the high-energy part of the PL spectrum of the blend, as the absorption of $\mathrm{PC_{70}BM}$ in this spectral range is lower compared to PCDTBT. Additionally the increased disorder in the intermixed donor--acceptor blends may also cause an increase of the relative height of the 0-0 transition. 
Due to the strongly reduced exciton lifetime in the blend, the dynamic redshift caused by the exciton migration to lower energetic sites has no significant influence on the spectral dynamics in the blend PL.

\begin{figure}
		\centering
 		\includegraphics[width=8.5cm]{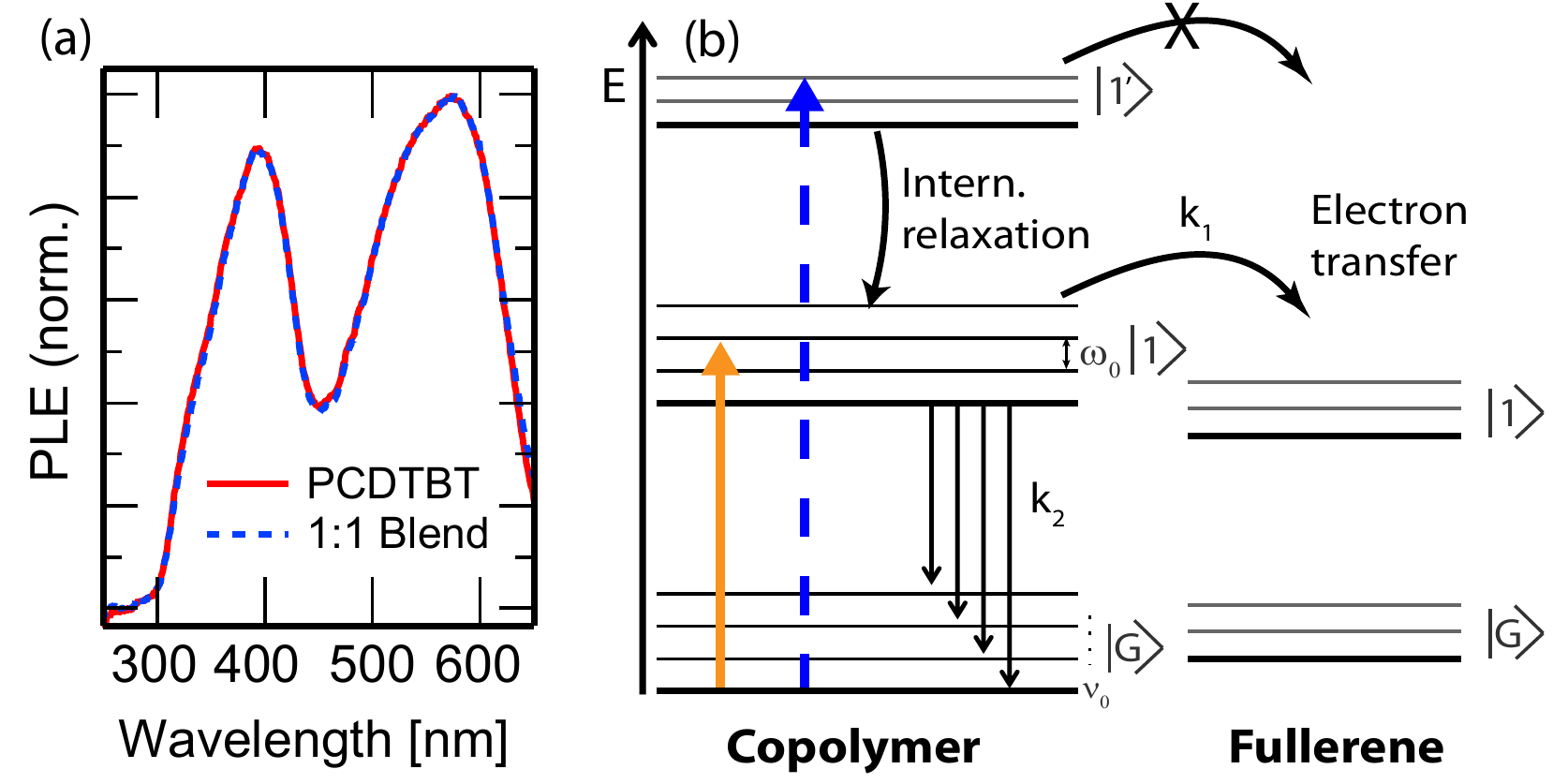}
		\caption{(a) PLE spectra of PCDTBT and a 1:1 PCDTBT:$\mathrm{PC_{70}BM}$ films. No indication of a competing relaxation channel following the excitation into the HEB can be assumed. (b) Energy scheme displaying the energy relaxation pathways in the copolymer-fullerene blends. Solid and dashed arrows indicate excitation of the LEB and HEB, respectively. The rates of electron transfer and the radiative decay to the ground state obey the inequality $k_{1}>k_{2}$.}
	\label{fig:EnergyScheme}
\end{figure}

After excitation into the HEB of the PCDTBT copolymer, the direct interaction between donor and acceptor by means of the electron or energy transfer to the $\mathrm{PC_{70}BM}$ competing with the internal excitation transfer within the copolymer moiety has to be considered.
Energy transfer towards the acceptor has been observed upon photoexcitation of copolymers F8BT and IFBT blended with $\mathrm{PC_{60}BM}$ \cite{soon_2011,cook_2006}. However, these copolymers were only excited via the low-energy absorption band and the energy transfer process was followed by triplet generation on the acceptor via intersystem crossing. To clarify the energy dependence of the interplay between internal relaxation within copolymer and intermolecular charge transfer, we performed the PL excitation (PLE) measurements. The comparison of the PLE spectra measured on pristine PCDTBT and on 1:1 blend (Figure \ref{fig:EnergyScheme} (a)) shows no indication of a direct excitation transfer from the high energy excited state towards the acceptor, as the two spectra are almost identical. This finding implies that for electrons created close to the donor---acceptor interface, which are expected to have a significant contribution to the PL of the 1:1 blend due to intercalation in this system\cite{mcgehee_2011}, the electron transfer to $\mathrm{PC_{70}BM}$ does not act as a significant alternative decay channel from the HEB (excitons created at a certain distance from the interface will relax before reaching it). Therefore, the internal partial CT on the copolymer backbone is expected to precede the electron transfer to the fullerene. 

Our results allow us to draw the picture of the excitation dynamics in copolymer PCDTBT, which is schematically shown in Figure \ref{fig:EnergyScheme} (b). As a result of the coupling between the BTC and benzothiadiazole units, the copolymer absorption spectrum exhibits two prominent transitions: the HEB corresponding to the excitation of the electron rich BTC donor segment while the LEB is a charge transfer band due to interaction of BTC with the adjacent benzothiadiazole. In PCDTBT:$\mathrm{PC_{70}BM}$ blends, direct excitation of the LEB leads to efficient charge separation at the donor acceptor interface. The radiative decay to the ground state is bypassed by this competitive process. Excitation to HEB is followed by a fast and efficient charge transfer towards benzothiadiazole in the sub-picosecond regime. This internal relaxation is followed by the charge separation in the presence of an acceptor.  

 \subsection{SUMMARY}
In this work we carried out a detailed analysis of the steady state absorption and PL as well as of the solvatochromic shift on the donor--acceptor copolymer PCDTBT and its building blocks---carbazole and 4,7-di(thien-2-yl)-2,1,3-benzothiadiazole as well as of the newly synthesized PBT-carbazole which represents a PCDTBT backbone segment without the thiadiazole end groups. This allowed us to decipher the functions of the constituent parts and their role in the copolymer. We assign the two prominent absorption bands of the copolymer to the absorption of the 2,7-di(thien-2-yl)carbazole and to the charge transfer state formed by the latter and the adjacent benzothiadiazole units, respectively. This is in contrast to the recent work of Banerji et al. \cite{banerji_2012} where both bands have been associated with charge transfer. Photoexcitation to the high energy band is followed by the internal relaxation within PCDTBT via the ultrafast charge transfer to benzothiadiazole group prior the radiative decay to the ground state. The decay dynamics of PCDTBT PL are dominated by a dynamic redshift of the spectrum due to conformational relaxation and exciton migration to more ordered regions (low energy sites) within the copolymer film. This shift indicates a reduced crystallinity in PCDTBT compared to, for example, P3HT. The addition of acceptor $\mathrm{PC_{70}BM}$ results in an efficient quenching of the copolymer PL within the first 50 ps, which is due to electron transfer to the acceptor. Comparing PL excitation spectra of PCDTBT and 1:1 PCDTBT:$\mathrm{PC_{70}BM}$ thin films we find no indication for a direct electron transfer from the high-energy excited state of PCDTBT towards fullerene acceptor. Instead, the internal energetic relaxation within the copolymer is assumed to precede the charge transfer to the fullerene acceptor underlining the importance of the coupling between functional copolymer units for the photovoltaic performance of the blends. Combining complementary experiments performed in solid and liquid phases, we believe to offer a physical picture on how the excitation energy is being transformed in the low band gap copolymer as well as in its fullerene blends. We think that our comparative studies of PCDTBT, its building blocks, following the "box of bricks" principle, provides an experimental tool that might help to decipher the interactions within novel D--A systems. 

\subsection{ACKNOWLEDGEMENTS}
The current work is supported by the Deutsche Forschungsgemeinschaft, DFG under the contract INST 93/623-1 FUGG and the Federal Ministry of Education and Research, BMBF in the framework of the GREKOS project (Grant No. 03SF0356B). C.D. gratefully acknowledges the support of the Bavarian Academy of Sciences and Humanities. We thank M. Steeger and Prof. C. Lambert, Physical Organic Chemistry at the University of Würzburg, for experimental support and the steady state PLE measurements.

\newpage 
\bibliographystyle{unsrt}
\bibliography{PCDTBT}

\begin{thebibliography}{10}

\bibitem{deibel_2010}
Carsten Deibel and Vladimir Dyakonov.
\newblock Polymer--fullerene bulk heterojunction solar cells.
\newblock {\em Reports on Progress in Physics}, 73(9):096401, 2010.

\bibitem{green_2011}
Martin~A Green, Keith Emery, Yoshihiro Hishikawa, and Wilhelm Warta.
\newblock {\em Progress in Photovoltaics: Research and Applications},
  19(1):84--92, 2011.

\bibitem{wong_2012}
Wileen Wong~Kromhout.
\newblock Ucla engineers create tandem polymer solar cells that set record for
  energy-conversion, {UCLA} newsroom, february 13th, 2012, February.

\bibitem{blouin_2008}
Nicolas Blouin, Alexandre Michaud, David Gendron, Salem Wakim, Emily Blair,
  Rodica {Neagu-Plesu}, Michel Belletete, Gilles Durocher, Ye~Tao, and Mario
  Leclerc.
\newblock {\em Journal of the American Chemical Society}, 130(2):732--742,
  2008.

\bibitem{heeger_2010}
Alan~J. Heeger.
\newblock {\em Chemical Society Reviews}, 39(7):2354--2371, 2010.

\bibitem{svensson_2003}
M.~Svensson, F.~Zhang, O.~Ingan{\"a}s, and {M.R.} Andersson.
\newblock {\em Synthetic Metals}, 135-136(0):137--138, 2003.

\bibitem{zhang_2006}
F.~Zhang, W.~Mammo, L. M. Andersson, S.~Admassie, M. R. Andersson, and
  O.~Ingan{\"a}s.
\newblock {\em Advanced Materials}, 18(16):2169--2173, 2006.

\bibitem{beaupre_2011}
Serge Beaupr{\'e}, Michel Belletete, Gilles Durocher, and Mario Leclerc.
\newblock {\em Macromolecular Theory and Simulations}, 20(1):13--18, 2011.

\bibitem{blouin_2007}
N.~Blouin, A.~Michaud, and M.~Leclerc.
\newblock {\em Advanced Materials}, 19(17):2295--2300, 2007.

\bibitem{park_2009}
Sung~Heum Park, Anshuman Roy, Serge Beaupr{\'e}, Shinuk Cho, Nelson Coates,
  Ji~Sun Moon, Daniel Moses, Mario Leclerc, Kwanghee Lee, and Alan~J. Heeger.
\newblock {\em Nat Photon}, 3(5):297--302, 2009.

\bibitem{steirer_2011}
K.~Xerxes Steirer, Paul~F. Ndione, N.~Edwin Widjonarko, Matthew~T. Lloyd, Jens
  Meyer, Erin~L. Ratcliff, Antoine Kahn, Neal~R. Armstrong, Calvin~J. Curtis,
  David~S. Ginley, Joseph~J. Berry, and Dana~C. Olson.
\newblock {\em Advanced Energy Materials}, 1(5):813--820, 2011.

\bibitem{sunb_2011}
Yanming Sun, Christopher~J. Takacs, Sarah~R. Cowan, Jung~Hwa Seo, Xiong Gong,
  Anshuman Roy, and Alan~J. Heeger.
\newblock {\em Advanced Materials}, 23(19):2226--2230, 2011.

\bibitem{tong_2010}
Minghong Tong, Nelson~E. Coates, Daniel Moses, Alan~J. Heeger, Serge
  Beaupr{\'e}, and Mario Leclerc.
\newblock {\em Physical Review B}, 81(12):125210, 2010.

\bibitem{banerji_2012}
Natalie Banerji, Eric Gagnon, {Pierre-Yves} Morgantini, Sebastian Valouch,
  Ali~Reza Mohebbi, {Jung-Hwa} Seo, Mario Leclerc, and Alan~J. Heeger.
\newblock {\em J. Phys. Chem. C}, 2012.

\bibitem{jespersen_2004}
Kim~G. Jespersen, Wichard J.~D. Beenken, Yuri Zaushitsyn, Arkady Yartsev, Mats
  Andersson, To{\~n}u Pullerits, and Villy Sundstr{\"o}m.
\newblock {\em The Journal of Chemical Physics}, 121(24):12613, 2004.

\bibitem{zang_2004}
C.~Zang.
\newblock {US} patent b2004 0229925, 2004.

\bibitem{vanmullekom_1998}
H.~A.~M van Mullekom, J.~A. J.~M Vekemans, and E.~W Meijer.
\newblock {\em Chemistry - A European Journal}, 4(7):1235--1243, 1998.

\bibitem{stahl_2006}
Rainer Stahl, Christoph Lambert, Conrad Kaiser, R{\"u}diger Wortmann, and Ruth
  Jakober.
\newblock {\em Chemistry - A European Journal}, 12(8):2358--2370, 2006.

\bibitem{inagaki_1972}
Takashi Inagaki.
\newblock {\em The Journal of Chemical Physics}, 57(6):2526, 1972.

\bibitem{chappell_2003}
John Chappell, David~G. Lidzey, Paul~C. Jukes, Anthony~M. Higgins, Richard~L.
  Thompson, Stephen {O'Connor}, Ilaria Grizzi, Robert Fletcher, Jim {O'Brien},
  Mark Geoghegan, and Richard A.~L. Jones.
\newblock {\em Nature Materials}, 2(9):616--621, 2003.

\bibitem{kirova_2010}
Serguei Brazovskii and Natasha Kirova.
\newblock {\em Chemical Society Reviews}, 39(7):2453, 2010.

\bibitem{Salaneck_1988}
W.~R. Salaneck, O.~Ingan{\"a}s, B.~Th{\'e}mans, J.~O. Nilsson, B.~Sj{\"o}gren,
  {J.-E.} {\"O}sterholm, J.~L. Br{\'e}das, and S.~Svensson.
\newblock {\em The Journal of Chemical Physics}, 89(8):4613, 1988.

\bibitem{feng_2007}
D.~{-Q} Feng, A.~N Caruso, Y.~B Losovyj, D.~L Shulz, and P.~A Dowben.
\newblock {\em Polymer Engineering \& Science}, 47(9):1359--1364, 2007.

\bibitem{clark_2007}
Jenny Clark, Carlos Silva, Richard~H. Friend, and Frank~C. Spano.
\newblock {\em Physical Review Letters}, 98(20):206406, 2007.

\bibitem{morin_2002}
{Jean-Franc{c}ois} Morin and Mario Leclerc.
\newblock {\em Macromolecules}, 35(22):8413--8417, 2002.

\bibitem{polander_2011}
Lauren~E. Polander, Laxman Pandey, Stephen Barlow, Shree~Prakash Tiwari, Chad
  Risko, Bernard Kippelen, {Jean-Luc} Bredas, and Seth~R. Marder.
\newblock {\em J. Phys. Chem. C}, 115(46):23149--23163, 2011.

\bibitem{Laquai_2011}
Fabian Etzold, Ian~A. Howard, Ralf Mauer, Michael Meister, {Tae-Dong} Kim,
  {Kwang-Sup} Lee, Nam~Seob Baek, and Frédéric Laquai.
\newblock {\em Journal of the American Chemical Society}, 133(24):9469--9479,
  2011.

\bibitem{banerji_2010}
Natalie Banerji, Sarah Cowan, Mario Leclerc, Eric Vauthey, and Alan~J. Heeger.
\newblock {\em Journal of the American Chemical Society}, 132(49):17459--17470,
  2010.

\bibitem{soon_2011}
Ying~W. Soon, Tracey~M. Clarke, Weimin Zhang, Tiziano Agostinelli, James
  Kirkpatrick, Clare {Dyer-Smith}, Iain {McCulloch}, Jenny Nelson, and James~R.
  Durrant.
\newblock {\em Chemical Science}, 2(6):1111, 2011.

\bibitem{cook_2006}
Steffan Cook, Hideo Ohkita, James~R. Durrant, Youngkyoo Kim, Jessica~J.
  {Benson-Smith}, Jenny Nelson, and Donal D.~C. Bradley.
\newblock {\em Applied Physics Letters}, 89(10):101128, 2006.

\bibitem{mcgehee_2011}
Zach~M Beiley, Eric~T Hoke, Rodrigo Noriega, Javier Dacu{\~n}a, George~F
  Burkhard, Jonathan~A Bartelt, Alberto Salleo, Michael~F Toney, and Michael~D
  {McGehee}.
\newblock {\em Advanced Energy Materials}, 1(5):954--962, 2011.

\end{thebibliography}

\end{document}